\documentclass[12pt]{article}

\usepackage{amsmath}
\usepackage{amssymb}
\usepackage{amsfonts}
\usepackage[mathscr]{euscript}

\title{\bf Environment Induced Bipartite Entanglement}

\author{Fabio Benatti$^{a,b}$,
Alexandra M. Liguori$^{a,b}$, Adam Nagy$^{a,c}$\\
\small $^a$Dipartimento di Fisica Teorica, Universit\`a di Trieste,
Strada Costiera 11,\\
\small 34014 Trieste, Italy\\
\small $^b$Istituto Nazionale di Fisica Nucleare, Sezione di
Trieste, 34100 Trieste, Italy \\
\small $^c$Budapest University of Technology and Economics,\\
\small Budapest, Muegyetem rkp.3-9, Hungary}

\date{\null}

\begin{document}

\maketitle

\vskip 2cm

\begin{abstract}
\noindent Recently, a sufficient condition on the structure of the
Kossakowski-Lindblad master equation has been given such that the
generated reduced dynamics of two qubits results entangling for at
least one among their initial separable pure states. In this paper
we study to which extent this condition is also necessary. Further,
we find sufficient conditions for bath-mediated entanglement
generation in higher dimensional bipartite open quantum systems.
\end{abstract}

\section{}
In standard quantum mechanics the focus is mainly upon
\textit{closed} physical systems, i.e. systems which can be
considered isolated from the external environment and whose
reversible time-evolution is described by one-parameter groups of
unitary operators. On the other hand, when a system $S$ interacts
with an environment $E$, it must be considered as an \textit{open}
quantum system whose time-evolution is irreversible and exhibits
dissipative and noisy effects. A standard way of obtaining a
manageable dissipative time-evolution of the density matrix
$\varrho_t$ describing the state of $S$ at time $t$ is to construct
it as the solution of a Liouville-type Master equation $\partial_t
\varrho_t = \mathbf{L}[\varrho_t]$. This can be done by tracing away
the environment degrees of freedom~\cite{nakajima, zwanzig} and by
performing a Markovian approximation~\cite{spohn, ben-flore}, i.e.
by studying the evolution on a slow time-scale and neglecting fast
decaying memory effects. Then the irreversible reduced dynamics of
$S$ is described by one-parameter semigroups of linear maps obtained
by exponentiating the generator $\mathbf{L}$ of Lindblad
type~\cite{kossak1, kossak2}:
$\gamma_t = e^{t\mathbf{L}}$, $t \geq 0$, such that %$\gamma_{t+s} = \gamma_t
%\circ \gamma_s = \gamma_s \circ \gamma_t$ ($s,t \geq 0$) and
$\varrho_t \equiv \gamma_t[\varrho]$.

The typical effect of noise and dissipation on a system $S$ immersed
in a large environment $E$ is decoherence; however, in certain
specific situations, the environment $E$ may even build quantum
correlations between the subsystems which compose $S$. This
possibility depends on the form of the Kossakowski matrix that
characterizes the dissipative part of the generator $\mathbf{L}$.
In~\cite{BFP} an inequality was found, involving the entries of such
a matrix which, if fulfilled, is sufficient to ensure that a
specific initial separable pure state of two qubits gets entangled.

This inequality is basically derived by looking at first derivatives
of evolving mean values that involve the generator only and not its
powers. In this paper we show that, apart from marginal cases whose
control needs second or higher powers of the generator, this
inequality is also necessary for entangling two qubits via immersion
within a common environment. Further, we consider higher dimensional
bipartite systems composed of two $d$-level subsystems embedded in a
common environment and provide sets of inequalities involving the
entries of a higher rank Kossakowski matrix. It turns out that if at
least one of these inequalities is fulfilled, the two parties get
entangled by their reduced dynamics.

\section{}
In this section we will consider a system $S$ composed of two
initially separable qubits immersed in a common external bath $E$
with which they weakly interact, but not directly interacting
between each other (\cite{ben-flore}, pages 3124-3126); thus the
total Hamiltonian is $H_T = H_1 + H_2 + H_B + \lambda H_I$, where
$H_1$, $H_2$ and $H_B$ are Hamiltonians pertaining to the first and
second qubit, respectively the bath within which they are immersed,
while the interaction Hamiltonian is given by
$H_I=\sum_{i=1}^3\big((\sigma_i \otimes \mathbb{I})\otimes B_i^{(1)}
+ (\mathbb{I} \otimes \sigma_i)\otimes B_i^{(2)} \big)$ with
$\mathbb{I}$ the identity $2\times 2$ matrix and $B_i^{(a)}, \,
a=1,2, \, i=1,2,3,$ bath operators that describe the interaction
with the two qubits. In the following, we shall use the convenient
notation $\sigma^{(1)}_i:=\sigma_i\otimes\mathbb{I}$ and
$\sigma^{(2)}_i:=\mathbb{I}\otimes\sigma_i$. By means of standard
\textit{weak coupling limit} techniques, the reduced dynamics of $S$
is given by the Master equation~\cite{ben-flore}
\begin{equation}
\label{master-eq} \frac{\partial \varrho_t}{\partial t} =
\mathbf{L}_H[\varrho_t] + \mathbf{D}[\varrho_t] =
-i[H_{eff},\varrho_t] + \mathbf{D}[\varrho_t]
\end{equation}
where $H_{eff} = H^{(1)} + H^{(2)} + H^{(12)} $, with
$H^{(a)}=\sum_{i=1}^3 h_i^{(a)}\sigma^{(a)}_i$,
$h^{(a)}_i\in\mathbb{R}$, $a=1,2$, Hamiltonians of the two qubits
independently,
\begin{equation}
\label{interaction-H} H^{(12)}=\sum_{i,j=1}^3 h_{ij}^{(12)}(\sigma_i
\otimes \sigma_j)\ ,\quad h^{(12)}_{ij}\in\mathbb{R}\ ,
\end{equation}
a Hamiltonian term describing a bath-mediated interaction between
the qubits, while
\begin{eqnarray}
\label{dissipation} \mathbf{D}[\varrho(t)] = \sum_{i,j=1}^3
&\Bigl(&\hskip -.4cm A_{ij}
\Bigl[\sigma^{(1)}_j\,\varrho\,\sigma^{(1)}_i\, -\,
\frac{1}{2}\{\sigma^{(1)}_i \sigma^{(1)}_j\,,\,\varrho \} \Bigr]
\nonumber \\
&+& C_{ij}\Bigl[\sigma^{(2)}_j\,\varrho\,\sigma^{(2)}_i\, -\,
\frac{1}{2}\{\sigma^{(2)}_i \sigma^{(2)}_j\,,\,\varrho \} \Bigr]
\nonumber \\
&+& B_{ij} \Bigl[\sigma^{(1)}_j\,\varrho\,\sigma^{(2)}_i\, -\,
\frac{1}{2}\{\sigma^{(1)}_j \sigma^{(2)}_i\,,\,\varrho \} \Bigr]
\nonumber \\
&+& B_{ji}^* \Bigl[\sigma^{(2)}_j\,\varrho\,\sigma^{(1)}_i\,-\,
\frac{1}{2}\{\sigma^{(1)}_i \sigma^{(2)}_j\,,\,\varrho \} \Bigr]
\Bigr)
\end{eqnarray}
is a Kossakowski-Lindblad contribution describing dissipation and
noise. The $3\times 3$ matrices $A=A^\dagger$, $C=C^\dagger$ and $B$
form the so-called Kossakowski matrix
\begin{equation}
\label{Kossa} K= \begin{pmatrix}
                 A  & B \\

                 B^\dagger  & C
\end{pmatrix}
\end{equation}
whose coefficients come from the Fourier transform of the bath
correlation functions (see~\cite{ben-flore}). In order to guarantee
full physical consistency, namely that ${\rm id}\otimes \gamma_t$ be
positivity preserving on all states of the compound system $S+S_d$
for any inert ancilla $S_d$, $\gamma_t$ must be completely positive
and this is equivalent to $K$ being non-negative~\cite{gorini,
lindblad}.

It is natural to call the generated semigroup entangling if there
exist at least two vector states $|\psi\rangle$ and
$|\varphi\rangle$ of the two parties such that $\gamma_t[Q]$ is
entangled for some $t>0$, where $Q:=|\psi\rangle\langle\psi|\otimes
|\varphi\rangle\langle\varphi|$\footnote{Notice that if $\gamma_t$
cannot entangle initially separable pure states then it cannot
entangle separable mixed states.}.

Because of the Peres-Horodecki criterion~\cite{horodecki}, in the
case of two qubits the semigroup $\gamma_t$ is entangling if and
only if there exist such a $Q$ and $t>0$ such that $T^{(2)} \circ
\gamma_t[Q]$ is not positive definite, where $T^{(2)} \equiv ({\rm
id} \otimes T):M_4(\mathbb{C})\mapsto M_4(\mathbb{C})$ is the
partial transposition on the second qubit with ${\rm
id}:M_2(\mathbb{C})\mapsto M_2(\mathbb{C})$ the identity operation
on the first qubit and $T:M_2(\mathbb{C})\mapsto M_2(\mathbb{C})$ is
the transposition with respect to the basis of eigenvectors of the
Pauli matrix $\sigma_3=\begin{pmatrix}1&0\cr0&-1\end{pmatrix}$.

Since $T\circ T={\rm id}$, the maps $\tilde{\gamma}_t := T^{(2)}
\circ \gamma_t \circ T^{(2)}$ form a semigroup, namely
\begin{equation}
\label{sem-comp-law}
\tilde{\gamma}_s\circ\tilde{\gamma}_t=\tilde{\gamma}_t\circ\tilde{\gamma}_s
=\tilde{\gamma}_{s+t}\ ,\qquad\forall\,s,t\geq0\ ,
\end{equation}
with generator $\tilde{\mathbf{L}} := T^{(2)} \circ \mathbf{L} \circ
T^{(2)}$. Setting
$\tilde{Q}:=T^{(2)}[Q]=|\psi\rangle\langle\psi|\otimes
|\varphi^*\rangle\langle\varphi^*|$, where $\vert\varphi^*\rangle$
denotes the vector state whose components in the chosen basis are
the conjugates of those of $\vert\varphi\rangle$, $\gamma_t$ results
entangling if and only if there exist a $Q$ and $t>0$ such that
$\tilde{\gamma}_t[\tilde{Q}] \ngeq 0$.

\bigskip

\noindent \textbf{Proposition 1}\quad The semigroup
$\gamma_t:M_4(\mathbb{C})\mapsto M_4(\mathbb{C})$ is entangling if
there exist a separable initial projector $Q\in M_4(\mathbb{C})$ and
a vector $\Phi \in \mathbb{C}^4$ such that:
\begin{equation}
\label{Q} \langle\Phi|\tilde{Q}|\Phi\rangle = 0 \qquad
\hbox{and}\qquad
\langle\Phi|\tilde{\mathbf{L}}[\tilde{Q}]|\Phi\rangle<0\ .
\end{equation}
Vice versa, the semigroup $\gamma_t$ cannot be entangling if
\begin{equation}
\label{L[Q]} \langle\Phi|\tilde{\mathbf{L}}[\tilde{Q}]|\Phi\rangle>0
\end{equation}
for all initial separable projectors $Q\in M_4(\mathbb{C})$ and
vectors $\Phi\in\mathbb{C}^4$ such that
$\langle\Phi|\tilde{Q}|\Phi\rangle = 0$.
\medskip

\noindent \textbf{Remark 1}\quad Notice that in case of an equality,
the argument of Proposition 1 cannot be used to conclude that
entanglement is or is not generated by the semigroup $\gamma_t$. A
concrete instance of this fact will be given in Example 1.

\medskip

\noindent \textbf{Proof:}\quad As stated above, $\gamma_t$ results
entangling if and only if there exist $Q$ and $t>0$ such that
$\tilde{\gamma}_t[\tilde{Q}] \ngeq 0$, i.e. if and only if there
exist $Q$, $t>0$ and $\Phi \in \mathbb{C}^4$ such that
$\langle\Phi|\tilde{\gamma}_t[\tilde{Q}]|\Phi\rangle < 0$. Since
$\langle\Phi|\tilde{Q}|\Phi\rangle \geq0$, the latter condition is
equivalent to the existence of a smallest $t^*\geq 0$ and
$\epsilon_0>0$ such that
$$
\langle\Phi|\tilde{\gamma}_{t^*}[\tilde{Q}]|\Phi\rangle = 0\quad (a)
\qquad\hbox{and} \qquad
\langle\Phi|\tilde{\gamma}_{t^*+\epsilon}[\tilde{Q}]|\Phi\rangle
<0\quad \forall \epsilon_0\geq\epsilon>0\qquad (b)\ .
$$
The assumption on $t^*$ means that $\tilde{\gamma}_{t^*}[\tilde{Q}]$
is still separable, whence it can be decomposed in a convex sum of
pure separable projectors, $\tilde{\gamma}_{t^*}[\tilde{Q}] =
\sum_{i,j}\lambda_{ij}Q_{ij}$, $0\leq\lambda_{ij}\leq1$. Then,
condition $(a)$ implies $\langle\Phi|Q_{ij}|\Phi\rangle = 0$ for all
$Q_{ij}$, while, from the semigroup composition
law~(\ref{sem-comp-law}) and condition (b) it follows that there
exist $i$ and $j$ such that
$\langle\Phi|\tilde{\gamma}_{\epsilon}[Q_{ij}]|\Phi\rangle<0$ for
all $0<\epsilon\leq\epsilon_0$. The continuity of the semigroup
formed by the $\tilde{\gamma}_t$ implies that (see for
instance~\cite{bratteli})
$$
\lim_{\epsilon \to
0^+}\langle\Phi|\frac{\tilde{\gamma}_{\epsilon}[Q_{ij}]-Q_{ij}}{\epsilon}
|\Phi\rangle=\langle\Phi|\tilde{\mathbf{L}}[Q_{ij}]|\Phi\rangle
=\lim_{\epsilon \to
0^+}\frac{\langle\Phi|\tilde{\gamma}_\epsilon[Q_{ij}]|\Phi\rangle}{\epsilon}
\ .
$$
If $\gamma_t$ is entangling, then for at least one of the $Q_{ij}$
it must be true that $\langle\Phi|Q_{ij}|\Phi\rangle=0$ and
$\langle\Phi|\tilde{\mathbf{L}}[Q_{ij}]|\Phi\rangle\leq0$, so that
if~(\ref{L[Q]}) holds as stated, $\gamma_t$ cannot be entangling.
Vice versa, if~(\ref{Q}) holds then
$\tilde{\gamma}_t[\tilde{Q}]\ngeq 0$ in a right neighborhood of
$t=0$ and the semigroup $\gamma_t$ is entangling.\\
\phantom{h}\hfill $\Box$
\medskip

In order to concretely apply the previous result, we shall introduce
the following notations. For given
$|\psi\rangle\,,\,|\varphi\rangle\in\mathbb{C}^2$, let $|u\rangle$,
$|v\rangle$ denote the vectors in $\mathbb{C}^3$ with components
\begin{equation}
\label{u-v} u_i:=\langle \psi|\sigma_i|\psi_\bot\rangle \, , \quad
v_i:=\epsilon_i\langle \varphi^*|\sigma_i|\varphi_\bot^*\rangle=
\langle\varphi_\bot\vert\sigma_i\vert\varphi\rangle\, ,
\end{equation}
where $\sigma_i$, $i=1,2,3$ are the Pauli matrices in the chosen
standard representation, whence, under transposition,
$\sigma_i^T=\epsilon_i\sigma_i$, with $\epsilon_i=+1$ when $i=1,3$
and $\epsilon_i=-1$ when $i=2$. Moreover, $\{\psi,\psi_\bot\}$,
$\{\varphi,\varphi_\bot\}$ are the orthonormal bases in
$\mathbb{C}^2$ corresponding to $\psi$ and $\varphi$. Let $C^T$
denote the transposition of the $3\times 3$ matrix $C$
in~(\ref{Kossa}), $Re(B)$ the $3\times 3$ matrix whose entries are
$\displaystyle Re(B)_{ij}:=\frac{B_{ij}+B^*_{ij}}{2}$ and $h^{(12)}$
is the $3\times 3$ real matrix formed by the coefficients
$h^{(12)}_{ij}$ of $H^{(12)}$ in~(\ref{interaction-H}).

From Proposition 1, it follows that we have to focus on
$\tilde{\mathbf{L}}[\tilde{Q}]$. When $\mathbf{L} = \mathbf{L}_H +
\mathbf{D}$ as in~(\ref{interaction-H}) and in~(\ref{dissipation}),
the action of the new generator explicitly reads
\begin{eqnarray}
\nonumber \tilde{\mathbf{L}}[\varrho] &:=& T^{(2)} \circ \mathbf{L}
\circ T^{(2)}[\varrho] =
-i\sum_{i=1}^3\Bigl[h_i^{(1)}\,\sigma^{(1)}_i\,+\,
h_i^{(2)}\,\epsilon_i\,\sigma_i^{(2)}\,,\,\varrho \Bigr]\\
\nonumber &+& \sum_{i,j=1}^3\Bigl(ih_{ij}^{(12)}\,\epsilon_j\,
\sigma^{(2)}_j\,\varrho\,\sigma^{(1)}_i
-ih_{ij}^{(12)}\,\epsilon_j\,\sigma^{(1)}_i\,\varrho\,
\sigma^{(2)}_j \Bigr) \\
\nonumber &+&
\sum_{i,j=1}^3A_{ij}\Bigl(\sigma^{(1)}_j\,\varrho\,\sigma^{(1)}_i\,
-\,
\frac{1}{2}\,\Bigl\{\sigma^{(1)}_i\sigma^{(1)}_j\,,\,\varrho\Bigr\}\Bigr)\\
&+&
\sum_{i,j=1}^3C_{ji}\epsilon_i\epsilon_j\,\Bigl(\sigma^{(2)}_j\,\varrho\,
\sigma^{(2)}_i\, -\, \frac{1}{2}\,\Bigl\{\sigma^{(2)}_i
\sigma^{(2)}_j\,,\,\varrho\Bigr\} \Bigr)
\nonumber \\
&+& \sum_{i,j=1}^3B_{ij}\epsilon_i
\Bigl(\sigma^{(1)}_j\sigma^{(2)}_i\,\varrho\,
-\,\frac{1}{2}\,\sigma^{(1)}_j\,\varrho\,\sigma^{(2)}_i \,
-\,\frac{1}{2}\,\sigma^{(2)}_i\,\varrho\,\sigma^{(1)}_j \Bigr)
\nonumber \\
&+& \sum_{i,j=1}^3 B_{ji}^* \epsilon_j \Bigl(\varrho\,\sigma^{(1)}_i
\sigma^{(2)}_j\, -\,\frac{1}{2}\,\sigma^{(1)}_i\,\varrho\,
\sigma^{(2)}_j\,-\,
\frac{1}{2}\,\sigma^{(2)}_j\,\varrho\,\sigma^{(1)}_i\Bigr)\ .
\label{L-tilde}
\end{eqnarray}
By regrouping the terms in~(\ref{L-tilde}) as in(\ref{dissipation}),
one sees that, with respect to~(\ref{Kossa}), the Kossakowski matrix
associated with $\tilde{\mathbf{L}}$ is now
\begin{equation}
\label{tilde-Kossa} \tilde{K}=\begin{pmatrix}
A&\tilde{B}\\
\tilde{B}^\dagger&\tilde{C}
\end{pmatrix}\ ,\quad
\tilde{B}_{ij}:=-\varepsilon_i\Bigl( \frac{B_{ij}+B^*_{ij}}{2}\,
+\,i\,h^{(12)}_{ji}\Bigr)\ ,\quad
\tilde{C}_{ij}=\varepsilon_i\varepsilon_j\,C_{ji}\ .
\end{equation}

Notice that, in spite of the fact that~(\ref{Kossa}) is positive
semi-definite, $\tilde{K}$ need not be so and therefore
$\tilde{\gamma}_t$ is not necessarily completely positive or even
positive; this allows for the possibility that
$\tilde{\gamma}_t[\tilde{Q}]$ be not positive semi-definite.

Proposition 1 tells us that we can concentrate on the mean values of
$\tilde{\mathbf{L}}[\tilde{Q}]$ with respect to
$\Phi\in\mathbb{C}^4$ that belong to the subspace orthogonal to
$\tilde{Q}$. Therefore, we can restrict our attention upon the
matrix $\tilde{Q}^\bot\tilde{\mathbf{L}}[\tilde{Q}]\tilde{Q}^\bot$,
$\tilde{Q}^\bot:=\mathbb{I}-\tilde{Q}$, that we will represent with
respect to the following orthonormal basis
\begin{eqnarray}
\label{basis}|\Psi_1\rangle:=|\psi\rangle \otimes |\varphi^*\rangle
\, , \qquad |\Psi_2\rangle:=|\psi\rangle \otimes
|\varphi_{\bot}^*\rangle \, , \nonumber \\
|\Psi_3\rangle:=|\psi_{\bot}\rangle \otimes |\varphi^*\rangle \, ,
\qquad |\Psi_4\rangle:=|\psi_{\bot}\rangle \otimes
|\varphi_{\bot}^*\rangle \, ,
\end{eqnarray}
where $|\psi\rangle$, $|\varphi\rangle$ are the $2$-dimensional
vectors which define $Q$, and
$\tilde{Q}=|\Psi_1\rangle\langle\Psi_1|$.

In calculating the matrix elements
$\langle\Psi_i\vert\tilde{\mathbf{L}}[\tilde{Q}]\vert\Psi_j\rangle$,
one observes that only two scalar products contribute to them,
either of the form $\langle\Psi_i\vert
\mathbb{I}\otimes\sigma_j\vert\Psi_1\rangle$ or of the form
$\langle\Psi_i\vert \sigma_j\otimes \mathbb{I}\vert\Psi_1\rangle$.
So $\textbf{L}^\perp:=\tilde{Q}^\bot
\tilde{\mathbf{L}}[\tilde{Q}]\tilde{Q}^\bot =
\begin{pmatrix}
                0 &   0  & 0 & 0 \\

                0 &  M_{22}  & M_{23} & 0 \\

                0 &  M_{23}^*  & M_{33} & 0 \\

                0 &  0  & 0 & 0
\end{pmatrix} \ ,
$ where
$M_{ij}:=\langle\Psi_i|\tilde{\mathbf{L}}[\tilde{Q}]|\Psi_j\rangle$;
explicitly,
\begin{eqnarray}
\label{M22} M_{22}&=& \sum_{i,j}C_{ji}\epsilon_i\epsilon_j\langle
\varphi_\bot^*|\sigma_j|\varphi^*\rangle \langle
\varphi^*|\sigma_i|\varphi_\bot^*\rangle = \langle v|C^T|v\rangle
\\
\label{M33} M_{33}&=& \sum_{i,j}A_{ij}\langle \psi_\bot|\sigma_j
|\psi\rangle \langle \psi|\sigma_i|\psi_\bot\rangle = \langle
u|A|u\rangle
\\
\nonumber M_{23}&=& -\sum_{i,j}\Big(ih_{ij}^{(12)} + \frac{B_{ji}+
B^*_{ji}}{2}\Big)\epsilon_j\langle
\varphi_\bot^*|\sigma_j|\varphi^*\rangle\langle
\psi|\sigma_i|\psi_\bot\rangle\\
\label{M23} &=& -\langle v|\Bigl(i(h^{(12)})^T +
Re(B)\Bigr)|u\rangle\ . \;
\end{eqnarray}
\medskip

\noindent \textbf{Proposition 2} \quad Given an initial projector
$M_4(\mathbb{C})\ni
Q=|\psi\rangle\langle\psi|\otimes|\varphi\rangle\langle\varphi|$,
with $\vert\psi\rangle\, ,\, \vert\varphi\rangle\in\mathbb{C}^2$,
consider the matrix $M=\begin{pmatrix}M_{22}&M_{23}\cr
M_{23}^*&M_{33}\end{pmatrix}$ with entries as
in~(\ref{M22})--(\ref{M23}). Then,
\begin{enumerate}
\item
if $Det(M) \equiv \langle u|A|u\rangle \langle v|C^T|v\rangle -
|\langle v|\Bigl(Re(B) + i(h^{(12)})^T\Bigr)|u\rangle|^2 < 0$, for
at least one pair $\vert\psi\rangle\, ,\,
\vert\varphi\rangle\in\mathbb{C}^2$, the semigroup $\gamma_t$ with
generator as
in~(\ref{master-eq}),~(\ref{interaction-H}),~(\ref{dissipation})
entangles $Q$;
\item
if $Det(M) \equiv \langle u|A|u\rangle \langle v|C^T|v\rangle -
|\langle v|\Bigl(Re(B) + i(h^{(12)})^T\Bigr)|u\rangle|^2 > 0$, for
all choices of
$\vert\psi\rangle\,,\,\vert\varphi\rangle\in\mathbb{C}^2$, the
semigroup $\gamma_t$ is not entangling.
\end{enumerate}
\medskip

\noindent \textbf{Proof:}\quad The proof of the first statement is a
simple application of Proposition 1. If $Det(M) < 0$, there exists a
vector $|\Phi\rangle$ such that $\tilde{Q}|\Phi\rangle =0$ and
$\langle\Phi|\tilde{\mathbf{L}}[\tilde{Q}]|\Phi\rangle<0$. Then a
first order  expansion in $t\geq 0$ gives
$\langle\Phi|\tilde{\gamma}_t[\tilde{Q}]|\Phi\rangle \simeq
t\langle\Phi|\tilde{\mathbf{L}}[\tilde{Q}]|\Phi\rangle<0$. This
implies that $\tilde{\gamma}_t[\tilde{Q}]$ is not positive
semi-definite in a right neighborhood of $t=0$, whence $\gamma_t[Q]$
is entangled~\footnote{This is essentially what was proved
in~\cite{BFP}.}. \\

 If $Det(M)>0$, then~(\ref{L[Q]}) holds so, if
$\tilde{Q}|\Phi\rangle=0$ and $\Phi\neq\Psi_4$ (and $\Phi\neq\Psi_1$
since we only need to consider the subspace orthogonal to
$\tilde{Q}=|\Psi_1\rangle\langle\Psi_1|$), a first order expansion
in $t\geq0$ gives
$\langle\Phi|\tilde{\gamma}_t[\tilde{Q}]|\Phi\rangle \simeq
t\langle\Phi|\tilde{\mathbf{L}}[\tilde{Q}]|\Phi\rangle =
t\langle\Phi|\textbf{L}^\perp|\Phi\rangle>0$.

If $\Phi=\Psi_4$, then $\langle\Phi|\textbf{L}^\perp|\Phi\rangle=0$
and the argument based on the first derivative seems to be not
conclusive; however, $\Psi_4$ is separable and therefore
$\langle\Psi_4|\tilde{\gamma}_t[\tilde{Q}]|\Psi_4\rangle\geq0$ for
all $t\geq0$. Hence, if $Det(M)>0$ for all choices of
$\vert\psi\rangle\, ,\, \vert\varphi\rangle\in\mathbb{C}^2$,
$\gamma_t$ cannot be entangling. \hfill$\Box$
\medskip

Unlike the last part of the previous proof, when $M$ has an
eigenvalue equal to zero, namely if $\langle u|A|u\rangle \langle
v|C^T|v\rangle= |\langle v|\Bigl( Re(B) +
i(h^{(12)})^T\Bigr)|u\rangle|^2$, then, in order to check whether
the semigroup $\gamma_t$ is entangling or not, one has to go to the
second or higher order terms in the small $t\geq0$ expansion of
$\langle\Psi|\tilde{\gamma}_t[\tilde{Q}]|\Psi\rangle$. In fact, if
$Det(M)=0$, there exists $|\Psi_1^\bot\rangle$ such that
$\textbf{L}^\perp|\Psi_1^\bot\rangle = 0$ so that
$\langle\Psi_1^\bot|\tilde{\gamma}_t[\tilde{Q}]|\Psi_1^\bot\rangle
\simeq
t^2/2\langle\Psi_1^\bot|\tilde{\mathbf{L}}^2[\tilde{Q}]|\Psi_1^\bot\rangle$.
As the following example shows, the non-negativity of the matrix $M$
does not fix the non-entangling character of $\gamma_t$: strict
positivity as in point 2. of Proposition 2 is necessary for this to
be true.
\medskip

\noindent \textbf{Example 1}\quad For sake of simplicity, we will
set to zero the Hamiltonian terms in~(\ref{master-eq}) and consider
a Kossakowski matrix of the form
$$
K=
\begin{pmatrix}
                A  & A  \\

                A  & A
\end{pmatrix} \quad \hbox{with}\quad
A= \begin{pmatrix}
                1  & 0 & i \\

                0  & 0 & 0 \\

                -i  & 0 & x
\end{pmatrix} \ ,
$$
with $x\geq1$ so that $K\geq0$. The purely dissipative generator
$$
\mathbf{L}[\rho]=\sum_{i,j=1}^3\sum_{p,q=1}^2A_{ij}\Bigl(\sigma_j^{(p)}\,\rho\,
\sigma_i^{(q)}-\frac{1}{2}\left\{\sigma_i^{(q)}\sigma_j^{(p)}\,,\,\rho\right\}
\Bigr)\ ,
$$
generates a continuous one-parameter semigroup of completely
positive maps $\gamma_t={\rm e}^{t\mathbf{L}}$. The Kossakowski
matrix~(\ref{tilde-Kossa}) associated with the generator
$\tilde{\mathbf{L}}$ of $\tilde{\gamma}_t:=T^{(2)}\circ\gamma_t\circ
T^{(2)}$ reads
$$
\tilde{K}=\begin{pmatrix}A&-Re(A)\cr -Re(A)^\dagger&A^T
\end{pmatrix}=\begin{pmatrix}
1&0&i&-1&0&0\cr 0&0&0&0&0&0\cr -i&0&x&0&0&-x\cr -1&0&0&1&0&-i\cr
0&0&0&0&0&0\cr 0&0&-x&i&0&x
\end{pmatrix}
$$
and is not positive definite, its non-zero eigenvalues being
$1\pm\sqrt{2}$ and $x\pm\sqrt{1+x^2}$. This ensures that
$\tilde{\gamma}_t$ is not completely positive; moreover, it turns
out that, for some values of $x$, it is not even positivity
preserving, leaving a chance that for some initial separable
projector $Q$ there exists $t>0$ such that
$\tilde{\gamma}_t[\tilde{Q}]$ might not be positive semi-definite.

Indeed, let $Q=|0\rangle\langle0|\otimes |0\rangle\langle0|=
\tilde{Q}$ where $\sigma_3|0\rangle=|0\rangle$ and
$\sigma_3|1\rangle=-|1\rangle$; then, $|u\rangle
=|v\rangle=(1,-i,0)$, and $\langle u|A|u\rangle = \langle
u|A^T|u\rangle = \langle u|Re(A)|u\rangle = 1$. Therefore,
$Det(M)=0$ and the the vector $|\Psi_1^\bot\rangle = |\Psi_2\rangle
+ |\Psi_3\rangle$ is eigenvector of
$\begin{pmatrix}0&0&0&0\cr0&1&-1&0\cr0&-1&1&0\cr0&0&0&0\end{pmatrix}$
with eigenvalue $0$. Considering the second order term in the
expansion of
$\langle\Psi_1^\bot|\tilde{\gamma}_t[\tilde{Q}]|\Psi_1^\bot\rangle$,
explicit calculations give $\langle
\Psi_1^\bot|\tilde{\mathbf{L}}^2[\tilde{Q}]|\Psi_1^\bot\rangle = 16x
- 24$; thus, the semigroup $\tilde{\gamma}_t$ is not positivity
preserving and $\gamma_t={\rm e}^{t\mathbf{L}}$ entangles $Q$ for
$1\leq x<3/2$, it does not do so for $x>3/2$, while if $x=3/2$ one
has to consider the third order term of the expansion in $t\geq0$.
\bigskip

\noindent \textbf{Remark 2}\quad If $A = B= C =0$ in~(\ref{Kossa})
and $H^{(a)}=0$ in~(\ref{master-eq}), the time-evolution is purely
reversible and governed by the interaction Hamiltonian
$H^{(12)}=\sum_{i,j=1}^3 h_{ij}^{(12)}(\sigma_i \otimes \sigma_j) $
in~(\ref{interaction-H}); then the sufficient condition for
entanglement in Proposition 2 reduces to (see in
particular~(\ref{M23}))
\begin{equation}
\label{ent-cond-H} |\langle u|h^{(12)}|v\rangle|^2 > 0
\end{equation}
for some $|u\rangle, |v\rangle\in\mathbb{C}^3$ of the
form~(\ref{u-v}).

In~\cite{dur-cirac} it is shown that there can be found local
unitary transformations
\begin{equation}
\label{orth-matrix} \sigma^A_i:=U_A^\dagger \sigma_i U_A =
\sum_{k=1}^3 O_{ik}^A\sigma_k \ ,\quad \sigma^B_j:=U_B^\dagger
\sigma_j U_B = \sum_{l=1}^3 O_{lj}^B\sigma_l \; ,
\end{equation}
such that $H^{(12)}=\sum_{i,j=1}^3h_{ij}^{(12)}\,
(\sigma_i\otimes\sigma_j)$ can always be recast as
$$
(U_A \otimes U_B)H^{(12)}(U_A^\dagger \otimes U_B^\dagger) :=
\hat{H}^{\pm} = \mu_1\sigma_1^A \otimes \sigma_1^B \pm
\mu_2\sigma_2^A \otimes \sigma_2^B + \mu_3\sigma_3^A \otimes
\sigma_3^B
$$
($\hat{H}^+$ if $Det(h^{(12)})\geq0$, $\hat{H}^-$ if
$Det(h^{(12)})<0$), where $\mu_1 \geq \mu_2 \geq \mu_3\geq0$ are the
sorted eigenvalues of $\sqrt{(h^{(12)})^\dagger h^{(12)}}$. Further,
the maximal entangling capability $\eta_{max}$ of an interaction
Hamiltonian $H^{(12)}$ is defined as
$$
\eta_{max} := max|\langle\chi^1 \otimes \chi^2|H^{(12)}|\chi^1_\bot
\otimes \chi^2_\bot\rangle| =
max|\langle\widetilde{\chi}^1\otimes\widetilde{\chi}^2|
\hat{H}^{\pm}|\widetilde{\chi}^1_\bot \otimes
\widetilde{\chi}^2_\bot\rangle|=(\mu_1+\mu_2)^2\ ,
$$
with $\vert\widetilde{\chi}^1\rangle:=U_A\vert\chi^1\rangle$ and
$\vert\widetilde{\chi}^2\rangle:=U_B\vert\chi^2\rangle$, the maximum
value being attained at $|\tilde{\chi}^1\rangle = |0\rangle_A$
eigenstate of $\sigma_3^A$ and $|\tilde{\chi}^2\rangle =
|1\rangle_B$ eigenstate of $\sigma_3^B$. Now
consider~(\ref{ent-cond-H}) and observe that
$$
|\langle u|h^{(12)}|v\rangle|^2 = |\sum_{i,j}h_{ij}^{(12)}\epsilon_j
\langle\psi_\bot|\sigma_i|\psi\rangle
\langle\varphi^*|\sigma_j|\varphi^*_\bot\rangle|=(\mu_1+\mu_2)^2
$$
for $|\psi\rangle = U_A|0\rangle_A$ and $|\varphi^*\rangle =
U_B|0\rangle_B$. Therefore, the maximal entanglement capability of a
two qubit Hamiltonian coincides with the the largest possible value
that fulfils the sufficient condition ~(\ref{ent-cond-H}).

\noindent \textbf{Remark 3}\quad Since partial transposition provides an exhaustive
entanglement witness also in the case of a two-level system coupled to a
three-level system~\cite{horodecki}, similar arguments as those developed above
can be applied to derive necessary and sufficient conditions for entanglement
generation even in this case. The proofs both of necessity and of sufficiency
would be the direct generalization of that of the two-qubit case with Pauli
matrices acting on the first subsystem and, for instance, Gell-Mann matrices
acting on the second, as will be explained at the beginning of the next
subsection. A concrete physical example of entanglement conditions for a
spin-1/2 coupled to a spin-1 will be given in a subsequent paper.
In the following, we shall rather discuss bipartite systems consisting of two
$d$-level systems; this includes, for instance, bipartite systems of $n$
qubits each, which are a natural generalization of the system previously considered,
although positivity under partial transposition is not sufficient to exclude
entanglement.

\section{}
The argument of the proof of sufficiency in Proposition 2 can be
extended to higher dimensional bipartite systems consisting of two
$d$-dimensional subsystems. It must be noticed that when $d\geq3$,
no extension of condition~(\ref{L[Q]}) is possible for there can be
entangled states which remain positive under partial
transposition~\cite{horodeckip}, that is, $\gamma_t$ might result
entangling despite $\tilde{\gamma}_t$ being positive on initially
separable states. On the other hand, though, the larger $d$ gets,
the more sufficient conditions we can obtain for the generation of
entanglement.

Let $\{F_k\}_{k=0}^{d^2-1}$, $F_0:=I_d/\sqrt{d}$, be an orthonormal
set of $d\times d$ Hermitian matrices such that ${\rm
Tr}(F_iF_j)=\delta_{ij}$ and, under transposition,
$F^T_k=\eta_k\,F_k$, where $\eta_k=\pm1$. For instance, as $F_k
\,$'s we can take the generalized Gell-Mann
matrices~\cite{georgi} which satisfy this request. \\
For a bipartite system where the two parties consist of $n$ qubits,
one chooses the matrices $F_k$ as tensor products of $n$ Pauli
matrices, whence $\eta_k$ is the product of $n$ $\varepsilon_i$,
where $\sigma_i^T=\varepsilon_i\,\sigma_i$ as in the previous
section.

If we consider a system $S$ composed of two subsystems each of
finite dimension $d$ immersed in a common external bath $E$ with
which they weakly interact, but not directly interacting between
each other, we can generalize the $2$-dimensional Master
equation~(\ref{master-eq}) with the $F_k$ matrices defined above.
Thus the total Hamiltonian is $H_T = H_1 + H_2 + H_B + \lambda H_I$,
where $H_1$, $H_2$ and $H_B$ are Hamiltonians pertaining to the
first and second subsystem, respectively to the bath, while the
interaction Hamiltonian is given by
$$
H_I=\sum_{i=1}^{d^2-1}\sum_{a=1}^2F^{(a)}_i\otimes B_i^{(a)}\
,\qquad F^{(1)}_i:=F_i\otimes\mathbb{I}\ ,\quad
F^{(2)}:=\mathbb{I}\otimes F_i\ ,
$$
with $\mathbb{I}$ the $d\times d$ identity matrix and $B_i^{(a)}$
bath operators. Again by means of standard \textit{weak coupling
limit} techniques, the reduced dynamics of $S$ is given by the
Master equation
\begin{equation}
\label{master-eq-gen} \frac{\partial \varrho_t}{\partial t} =
\mathbf{L}_H[\varrho_t] + \mathbf{D}[\varrho_t] =
-i[H_{eff},\varrho_t] + \mathbf{D}[\varrho_t]
\end{equation}
where $H_{eff} = H^{(1)} + H^{(2)} + H^{(12)} $, with
$H^{(a)}=\sum_{i=1}^{d^2-1} h_i^{(a)}F^{(a)}_i $,
$h_i^{(a)}\in\mathbb{R}$, $a=1,2$, Hamiltonians of the two
subsystems independently,
\begin{equation}
\label{interaction-H-gen} H^{(12)}=\sum_{i,j=1}^{d^2-1}
h_{ij}^{(12)}(F_i \otimes F_j) \, , \quad h_{ij}^{(12)} \, \in
\mathbb{R} \, ,
\end{equation}
a Hamiltonian term describing a bath-mediated interaction between
the subsystems, while
\begin{eqnarray}
\label{gen-dissipation} \mathbf{D}[\varrho(t)] =
\sum_{i,j=1}^{d^2-1} &\Bigl(&\hskip -.4cm A_{ij}
\Bigl[F^{(1)}_j\,\varrho\,F^{(1)}_i\,-\, \frac{1}{2}\{F^{(1)}_i
F^{(1)}_j\,,\,\varrho \} \Bigr]
\nonumber \\
&+& C_{ij}\Bigl[ F^{(2)}_j\,\varrho\,F^{(2)}_i\, -\,
\frac{1}{2}\{F^{(2)}_i F^{(2)}_j\,,\,\varrho \} \Bigr]
\nonumber \\
&+& B_{ij} \Bigl[F^{(1)}_j\,\varrho\,F^{(2)}_i\, -\,
\frac{1}{2}\{F^{(1)}_jF^{(2)}_i\,,\,\varrho \} \Bigr]
\nonumber \\
&+& B_{ji}^* \Bigl[F^{(2)}_j\,\varrho\,F^{(1)}_i\, -\,
\frac{1}{2}\{F^{(1)}_iF^{(2)}_j\,,\,\varrho \} \Bigr]\Bigr)\ ,
\end{eqnarray}
is a Kossakowski-Lindblad contribution describing dissipation and
noise. $A=A^\dagger$, $C=C^\dagger$ and $B$ are $(d^2-1) \times
(d^2-1)$ matrices which define a $2(d^2-1)\times 2(d^2-1)$
Kossakowski matrix $K=
\begin{pmatrix}
                 A  & B \\

                 B^\dagger  & C
\end{pmatrix} $.
As in the $d=2$ case, in order to guarantee the complete positivity
of the dynamical map $\gamma_t = e^{t\mathbf{L}}$ and thus its full
physical consistency against coupling with generic ancillas and the
existence of entangled states, the Kossakowski matrix must be
positive semi-definite, $K\geq0$ ~\cite{gorini, lindblad}.

As in Proposition 1 and Proposition 2, let
$Q=|\psi\rangle\langle\psi|\otimes|\varphi\rangle\langle\varphi|$,
with $|\psi\rangle\,,\,|\varphi\rangle\in\mathbb{C}^d$, be an
initial separable projector of the two $d$-level systems. We shall
then consider the semigroup $\tilde{\gamma}_t =
e^{t\tilde{\mathbf{L}}}$ and its generator
$\tilde{\mathbf{L}}=T^{(2)}\circ\mathbf{L}\circ T^{(2)}$, where
$T^{(2)}$ is the partial transposition operated on the second
factor. The form of  $\tilde{\mathbf{L}}$ is the same as
in~(\ref{L-tilde}) and~(\ref{tilde-Kossa}) with the $F_k$ matrices
instead of the Pauli matrices and $\eta_i$ in place of
$\varepsilon_i$.

According to Proposition 1 and the proof of point 1 in Proposition
2, in order to find sufficient conditions for $\gamma_t$ to be
entangling, we just have to study when
$\tilde{Q}^\bot\tilde{\mathbf{L}}[\tilde{Q}]\tilde{Q}^\bot$ has a
negative eigenvalue, where $\tilde{Q}=T^{(2)}[Q]
=\vert\psi\rangle\langle\psi\vert\otimes
\vert\varphi^*\rangle\langle\varphi^*\vert$,
$\tilde{Q}^\bot=\mathbb{I}-\tilde{Q}$.

Let $\{|\psi_i\rangle\}_{i=1}^d$ and $\{|\varphi_i\rangle\}_{i=1}^d$
be two orthonormal bases for the two parties, with
$\vert\psi_1\rangle=\vert\psi\rangle$ and
$\vert\varphi_1\rangle=\vert\varphi^*\rangle$. A convenient
enumeration for the corresponding basis of the composite system is
as follows: $|\Psi_{d(k-1)+\ell}\rangle
:=|\psi_k\otimes\varphi_\ell\rangle$ for $k,\ell=1,2,\ldots,d$. Set
$i=d(k-1)+\ell$, $i=1,2,\ldots,d^2$; then,
$\tilde{Q}:=|\Psi_1\rangle \langle \Psi_1|$ and $\tilde{Q}^{\bot}:=
\sum_{i=2}^{d^2}|\Psi_i\rangle \langle \Psi_i|$.

Since $F^{(1)}_i=F_i\otimes\mathbb{I}$ and
$F^{(2)}_i=\mathbb{I}\otimes F_i$, with respect to the chosen basis,
only the entries $M_{ij}:=\langle\Psi_i\vert\tilde{Q}^\bot
\tilde{\mathbf{L}} [\tilde{Q}]\tilde{Q}^\bot\vert\Psi_j\rangle$ with
either $k=1$ or $\ell=1$ in $i=\ell+d(k-1)$ survive, while all those
with $k \neq 1$ and $\ell \neq 1$ vanish. There are $2(d-1)$ basis
vectors with either $k=1$ or $\ell=1$:
\begin{equation}
\label{vectors}
\begin{matrix}
& \vert\Psi_{2}\rangle=\vert\psi_1\otimes\varphi_2\rangle\ ,\
\vert\Psi_{3}\rangle=\vert\psi_1\otimes\varphi_3\rangle\ ,\ldots\,
\vert\Psi_{d}\rangle=\vert\psi_1\otimes\varphi_d\rangle \cr \cr &
\vert\Psi_{d+1}\rangle=\vert\psi_2\otimes\varphi_1\rangle\ ,\
\vert\Psi_{2d+1}\rangle=\vert\psi_3\otimes\varphi_1\rangle\
,\ldots\,
\vert\Psi_{(d-1)d+1}\rangle=\vert\psi_d\otimes\varphi_1\rangle\ ,
\end{matrix}
\end{equation}
and $(d-1)^2$ vectors with $k \neq 1$ and $\ell \neq 1$. Therefore
we can focus on $\tilde{Q}^\bot
\tilde{\mathbf{L}}[\tilde{Q}]\tilde{Q}^\bot$ restricted to the
subspace spanned by the vectors in~(\ref{vectors}), i.e. on a
$2(d-1)\times 2(d-1)$ non-zero submatrix that we will call $M$. This
matrix is composed of four $(d-1)$-dimensional square blocks and its
entries can be written in analogy to the $d=2$ case generalizing the
vectors $|u\rangle$, $|v\rangle$ in~(\ref{u-v}). Explicitly, we
define $(d-1)$ vectors $|u^{(n)}\rangle$ and $(d-1)$ vectors
$|v^{(m)}\rangle$ with $(d^2-1)$ components each, given by
\begin{eqnarray}
\label{u-v-gen}
u_i^{(n)}&:=&\langle \psi_1|F_i|\psi_n\rangle \, , \quad n=d+1,d+2,\ldots \, ,2d-1, \; i=1, \ldots \, ,d^2-1, \\
v_i^{(m)}&:=& \eta_i \langle \varphi_1|F_i|\varphi_m\rangle \, ,
\quad m=2,\ldots,d, \; i=1, \ldots \, ,d^2-1.
\end{eqnarray}
Further, we introduce a Hermitian $2(d-1)\times 2(d-1)$ matrix
$M=[M_{\alpha,\beta}]$ with entries
\begin{eqnarray*}
&& M_{\alpha,\beta}:=\langle
v^{(\alpha+1)}|C^T|v^{(\beta+1)}\rangle\, , \quad
\alpha,\beta=1,2,\ldots,d-1 \cr
&&\\
&& M_{\alpha,\beta}:=\langle u^{(\alpha+1)}|A|u^{(\beta+1)}\rangle\,
,\quad \alpha,\beta=d,d+1,\ldots,2(d-1)\cr
&&\\
&& M_{\alpha,\beta}:=-\langle
v^{(\alpha+1)}|\Bigl(i(h^{(12)})^T+Re(B)\Bigr)|u^{(\beta+1)}\rangle\,
, \quad \alpha=1,\ldots,d-1,\,
\beta=d,\ldots,2(d-1)\\
&&\\
&& M_{\alpha,\beta}:=-\overline{\langle
v^{(\beta+1)}|\Bigl(i(h^{(12)})^T+Re(B)\Bigr)
|u^{(\alpha+1)}\rangle}\, , \quad \alpha=d,\ldots,2(d-1), \,
\beta=1,\ldots,d-1\ .
\end{eqnarray*}
In Proposition 2, we saw that the negativity of the determinant of
the matrix $M=[M_{\alpha\beta}]$ with $d=2$ is a sufficient
condition for a bath-mediated entanglement of an initial separable
projector. For a bipartite system composed of two $d$-level
subsystems, the argument generalizes as follows.
\medskip

\noindent \textbf{Proposition 3}\quad If at least one of the
principal minors of the $2(d-1)\times 2(d-1)$ matrix
$M=[M_{\alpha\beta}]$ is negative, then the semigroup $\gamma_t$ is
entangling. Therefore there are $2^d(2^{d-2}-1)+1$ conditions at the
most, each one of them ensuring bipartite entanglement generation
through immersion in a common environment.
\medskip

\noindent \textbf{Proof:}\quad Let $R$ be one of $M$'s principal
sub-matrices. If $Det(R) < 0$, then there exists a vector
$|\Phi\rangle$ in the support of
$\tilde{Q}^\bot\tilde{\mathbf{L}}[\tilde{Q}]\tilde{Q}^\bot$ such
that $\langle\Phi|\tilde{\mathbf{L}}[\tilde{Q}]|\Phi\rangle <0$ and
$\langle\Phi|\tilde{Q}|\Phi\rangle=0$. Thus, an expansion at small
times $t\geq0$ yields
$$
\langle\Phi|\tilde{\gamma}_t[\tilde{Q}]|\Phi\rangle \simeq
t\langle\Phi|\tilde{\mathbf{L}}[\tilde{Q}]|\Phi\rangle < 0 \, ,
$$
which implies that $\tilde{\gamma}_t[\tilde{Q}]$ is not positive
semi-definite in a right neighborhood of $t=0$ and $\gamma_t[Q]$
becomes entangled in that time-interval.

Being a $2(d-1)\times2(d-1)$ matrix, $M$ has $2^{2(d-1)}-1$
principal sub-matrices; however,  since $A$ and $C$
in~(\ref{gen-dissipation}) are positive matrices, all their
$2(2^{d-1}-1)$ principal minors cannot be negative: therefore, all
the diagonal elements of $M$ are surely non-negative. Thus we are
left with, at most, $(2^{2(d-1)}-1) - 2(2^{d-1}-1)= 4^{d-1}-2^d +1$
principal minors that are not necessarily positive~\footnote{In the
two qubit case, $d=2$ and we get $4-2^2+1=1$ sufficient condition
for entanglement, as found in Proposition 2.}. \hfill$\Box$
\medskip

\noindent \textbf{Example 2}\quad We will consider four qubits
$(1,2,3,4)$ immersed in a dissipative environment such that their
states evolve in time according to a Master
equation~(\ref{master-eq-gen}) with a purely dissipative generator
of the form
\begin{eqnarray}
\nonumber \mathbf{L}[\varrho]&=&\sum_{p=1}^4 \sum_{i,j=1}^3
C_{ij}^{(1)} \Big( \sigma_j^{(p)}\,\varrho\,\sigma_i^{(p)} -
\frac{1}{2}
\{\sigma_i^{(p)}\!\sigma_j^{(p)}\,,\,\varrho\} \Big)\\
\label{4-qubit} &+& \sum_{p\neq q=1}^4 \sum_{i,j=1}^3 C_{ij}^{(2)}
\Big( \sigma_j^{(p)}\,\varrho\,\sigma_i^{(q)} - \frac{1}{2}
\{\sigma_i^{(p)}\!\sigma_j^{(q)}\,,\,\varrho\} \Big) \, .
\end{eqnarray}
In terms of the matrices $F_k$ in~(\ref{gen-dissipation}), one has
$F_k := F_{i+3(p-1)} = \frac{1}{\sqrt{2}}\sigma_i^{(p)}$, $i=1,2,3$,
$p=1,2$ and a Kossakowski matrix
$$
K_4 =
\begin{pmatrix}
                C^{(1)}  & C^{(2)}  & C^{(2)} & C^{(2)} \\

                C^{(2)}  & C^{(1)}  & C^{(2)} & C^{(2)} \\

                C^{(2)}  & C^{(2)}  & C^{(1)} & C^{(2)}  \\

                C^{(2)}  & C^{(2)}  & C^{(2)} & C^{(1)}  \\

\end{pmatrix} \, .
$$
Taking $C^{(1)} =
\begin{pmatrix}
                1  & iz  & 0  \\

                -iz  & 1  & 0  \\

                0  & 0  & 0   \\
\end{pmatrix}$
and $C^{(2)}  =
\begin{pmatrix}
                x  & 0  & 0  \\

                0  & -x  & 0  \\

                0  & 0  & 0   \\
\end{pmatrix}$
with $z,x$ real numbers, in order for the semigroup generated
by~(\ref{4-qubit}) to be completely positive, the Kossakowski matrix
$K_4$ must be positive semi-definite; this implies $z^2 + 9x^2 \leq
1$ as its non-zero eigenvalues are $1\pm\sqrt{x^2+z^2}$ and
$1\pm\sqrt{9x^2+z^2}$.

Consider the fully separable state $Q_4:=
\Bigl(|0\rangle\langle0|\otimes|0\rangle\langle0|\Bigr)\otimes
\Bigl(|0\rangle\langle0|\otimes|0\rangle\langle0|\Bigr)$,
$\sigma_3|0\rangle=|0\rangle$; the vectors defined
in~(\ref{u-v-gen}) are
$$
|u^{(5)}\rangle=
\begin{pmatrix}
                0  \\
                0  \\
                0   \\
                1   \\
                -i   \\
                0
\end{pmatrix} \ ,\quad
|v^{(2)}\rangle=
\begin{pmatrix}
                0  \\
                0  \\
                0   \\
                1   \\
                i   \\
                0
\end{pmatrix} \, ,\quad
|u^{(6)}\rangle=
\begin{pmatrix}
                1  \\
                -i  \\
                0   \\
                0   \\
                0   \\
                0
\end{pmatrix} \, ,\quad
|v^{(3)}\rangle=
\begin{pmatrix}
                1  \\
                i  \\
                0   \\
                0   \\
                0   \\
                0
\end{pmatrix} \, ,
$$
while $|u^{(7)}\rangle$ and $|v^{(4)}\rangle$ are the null vector.
Therefore, the $2(d-1)\times 2(d-1)=6\times 6$ matrix
$M=[M_{\alpha\beta}]$ reduces to a $4\times 4$ matrix of the form
$$
M = 2
\begin{pmatrix}
                1+z  & 0  & x & x \\

                0  & 1+z  & x & x \\

                x  & x  & 1+z & 0  \\

                x  & x  & 0 & 1+z  \\

\end{pmatrix}\, .
$$
Since $z^2\leq z^2+9x^2\leq1$, its principal minors of order $1$,
$2(1+z)$, are all positive; those of order $2$ are the determinants
of $ 2\,\begin{pmatrix} 1+z&0\cr 0&1+z
\end{pmatrix}\ ,\quad
2\,\begin{pmatrix} 1+z&x\cr x&1+z
\end{pmatrix}$,
and are also positive if $(1+z)^2>x^2$. Those of order $3$,
$D(x,z):=8(1+z)((1+z)^2-2x^2)$, namely the determinants of the
matrices
$$
2\,\begin{pmatrix} 1+z&0&x\cr 0&1+z&x\cr x&x&1+z
\end{pmatrix}\ ,\quad
2\,\begin{pmatrix} 1+z&x&x\cr x&1+z&0\cr x&0&1+z
\end{pmatrix}\ ,
$$
can nevertheless be negative. In fact, there is a region in the plane
$(x,z)$ where $z^2+9x^2 \leq 1$ and $(1+z)^2> x^2$, while
$(1+z)^2<2x^2$. The corresponding values $(x,z)$ ensure that $K_4$
is positive semi-definite, whereas $M$ is not; correspondingly, the
$4$-qubit dissipative dynamics entangles the two pairs $(1,2)$ and
$(3,4)$ initially in the pure separable state $Q_4$.

The fact that the principal minors of order $2$ are non-negative has
the following physical interpretation. The form of the generator of
the dissipative dynamics of the two pairs of qubits is such that if
we eliminate any pair of qubits by taking the trace
of~(\ref{4-qubit}) over their Hilbert spaces, a generator results
for the remaining pair $(i,j)$ of qubits that amounts to keeping
only the $i$-th and $j$-th row and column of $K_4$, thereby leading
to a same dissipative time-evolution associated with the Kossakowski
matrix $K_2=
\begin{pmatrix}
                C^{(1)}  & C^{(2)} \\

                C^{(2)}  & C^{(1)} \\
\end{pmatrix}$
for any pair $(i,j)$ of qubits. Moreover, given $Q_4$, any pair of
qubits is in the same state
$Q_2=|0\rangle\langle0|\otimes|0\rangle\langle0|$ with the vectors
$|u\rangle$ and $|v\rangle$ defined in~(\ref{u-v}) that read
$|u\rangle =(1,-i,0)$, $|v\rangle = (1,i,0)$ and yield a matrix $M=
2
\begin{pmatrix}
                1+z  & x \\

                x  & 1+z \\
\end{pmatrix}$.
By assumption $\hbox{Det}(M)>0$, thus, according to Proposition 2,
the two-qubit semigroup associated with the Kossakowski matrix $K_2$
cannot entangle $Q_2$; nonetheless, the four-qubit semigroup
associated with $K_4$ entangles $Q_4$.
\bigskip

\section{}
We have shown that the sufficient condition found in~\cite{BFP} for
the creation of entanglement between two qubits immersed in a common
environment by means of their reduced dissipative dynamics is, apart
from marginal cases, also necessary. Moreover, we have extended to
higher dimensional bipartite open quantum systems the basic argument
in~\cite{BFP} thereby obtaining sufficient conditions for entangling
a separable pure state in terms of the negativity of the principal
minors of certain matrices that depend on the generator of the
dynamics and on the given state to get entangled. Since the number
of principal minors increases with the dimension of the parties, for
more than two qubits a richer variety of noise-induced entanglement
is available. As an example, we provided a purely dissipative time
evolution that entangles two subsystems, each consisting of two
qubits, without entangling any two single qubits.
\bigskip

\noindent \textbf{Aknowledgement}\quad The authors want to thank R.
Floreanini for his valuable comments and suggestions.

\end{document}